\documentclass[aps,prb,reprint]{revtex4-1}
\usepackage{amsmath}
\usepackage{array}
\usepackage{amsfonts}
\usepackage{amssymb}
\usepackage{graphicx}
\usepackage{units,color}
\usepackage{hyperref}
\usepackage{epstopdf}
\usepackage{multirow}
\usepackage[utf8]{inputenc}

\providecommand{\U}[1]{\protect\rule{.1in}{.1in}}
\providecommand{\U}[1]{\protect\rule{.1in}{.1in}}
\providecommand{\U}[1]{\protect\rule{.1in}{.1in}}
\providecommand{\U}[1]{\protect\rule{.1in}{.1in}}
\begin{document}

\title{Equilibrium current vortices in rare-earth--doped simple metals}
\author{Adam B. Cahaya$^{1}$, Alejandro~O. Leon$^{2}$, Mojtaba Rahimi Aliabad%
$^{3}$, and Gerrit E.~W. Bauer$^{4,5,6}$}
\date{\today }

\begin{abstract}
Dilute alloys of rare earths have played a vital role in understanding
magnetic phenomena. Here, we model the ground state of dilute 4f rare-earth
impurities in light metals. When the 4f subshells are open (but not
half-filled), the spin-orbit coupling imprints a rotational charge current
of conduction electrons around rare-earth atoms. The sign and amplitude of
the current oscillate similar to the RKKY spin polarization. We compute the
observable effect, namely the \O rsted field generated by the current
vortices and the Knight shift.
\end{abstract}

\affiliation{$^1$ Department of Physics, Faculty of Mathematics and Natural
Sciences, Universitas Indonesia, Depok 16424, Indonesia} 
\affiliation{$^2$Departamento de F\'isica, Facultad de Ciencias Naturales,
Matem\'atica y del Medio Ambiente,
Universidad Tecnol\'ogica Metropolitana, Las Palmeras
3360, Ñuñoa 780-0003, Santiago, Chile} 
\affiliation{$^3$Nano-Structured
Coatings Institute of Yazd Payame Noor University, P.O. Code 89431-74559,
Yazd, Iran} 
\affiliation{$^5$ WPI-AIMR $\&$ CSRN, Tohoku University,
Sendai 980-8577, Japan} 
\affiliation{$^4$Institute for Materials Research, Tohoku University, Sendai
980-8577, Japan} 
\affiliation{$^6$Zernike Institute for Advanced
Materials, Groningen University, The Netherlands}
\maketitle
\section{Introduction}

The conversion of spin currents into excitations of the charge, phonon,
photon, or magnetization degrees of freedom and vice versa~\cite%
{OtaniSpinConv} often involves spin-orbit interactions (SOI)~\cite{SOT2013}.
Examples are the spin-orbit torques~\cite{Daichi2015}, charge pumping~\cite%
{ChargePumping2015}, magnetoelastic interactions~\cite{PhononMaekawa}, and
electric-field-induced magnetization dynamics~\cite{Elec1,Elec2}. The large
intraatomic SOI that governs the local moments of lanthanides with partially
filled 4f subshells causes novel spin charge coupling~\cite{voltage} and
affects device parameters such as the magnetic damping~\cite%
{REDoping,RECapping}. Rare-earth (RE) ions with local magnetic moments can
partially or entirely substitute the non-magnetic yttrium in the
ferrimagnetic insulator yttrium iron garnet Y$_{3}$Fe$_{5}$O$_{12}$ (YIG)~%
\cite{YIG}. The different magnetic sublattices of RE-IG strongly modify the
magnetic properties~\cite{RIG1,RIG2, RIG3}, causing, for example, different
compensation points for the magnetic and total angular moments~\cite%
{DoubleCompensation}. A more complex phenomenon is a double sign change of
the spin Seebeck effect~\cite{GIG}. Thulium iron garnet (Tm$_{3}$Fe$_{5}$O$%
_{12}$) films with perpendicular magnetization~\cite{Kubota2012,Tang2016,
TIG2018} can be switched by current-induced spin-orbit torques~\cite%
{TmIG0,TmIG1,TmIG2}.

These new developments come on top of decades of research on 4f electrons in
bulk metals~\cite{BookJensen}. For example, RE impurities in non-magnetic
metals cause an anomalous Hall effect at low doping concentrations~\cite%
{Fert}. The magnetization in rare-earth intermetallics originates from both
the 5d and 6s conduction electrons and 4f moments (see Ref.~\cite%
{REIntermetallics} and references therein). The hybridization of RE moments
with conduction electrons affects the susceptibility in rare-earth
dialuminides, REAl$_{2}$~\cite{REAl2}, or causes enhanced magnetic moments
of RE dopants in Ag and Au~\cite{Devine}.

In topological superconductors, such as the Fe(Te, Se), a vortex
with a Friedel-like oscillatory profile around magnetic impurities has been
reported~\cite{SCVortex}.

Here we present a theoretical study of the coupling of a 4f local moment
with the Fermi sea of a simple metal host. We predict a charge current
circling the impurity with a direction that oscillates radially, as
illustrated in Fig.~\ref{Fig0}. Our starting point for the interaction
between the local 4f moments and the conduction electrons is the Kondo
Hamiltonian~\cite{Kondo}. Its chirality induces a circulating current whose
vorticity is governed by the direction of the RE orbital moment and
generates an \O rsted magnetic field. The induced radial current
distribution oscillates with the same period as the RKKY spin polarization~%
\cite{RKKY} and for the same physical reason, i.e., the finite momentum
cut-off at the Fermi surface. The predicted trends should be observable {%
in principle by NMR\ or scanning microscopy. The present study
of the interaction between conduction electrons and 4f magnetic moments
contributes to understanding spintronic devices including rare-earth local
moments, such as interfaces between rare-earth iron garnets and nonmagnetic
metals.

\begin{figure}[b]
\includegraphics[width=5cm]{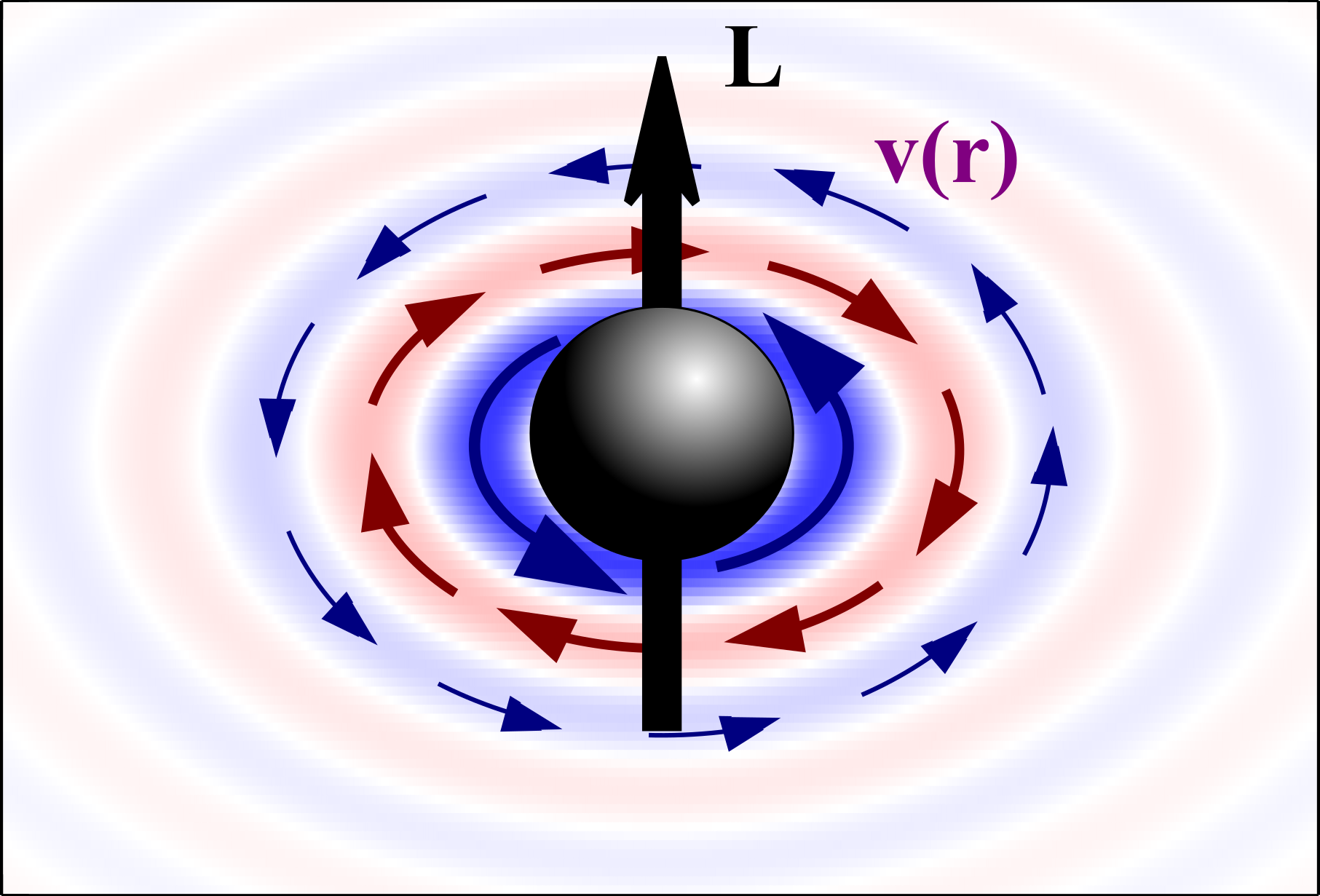}
\caption{Schematic representation of the rotational velocity field $\mathbf{v%
}(\mathbf{r})$ around a rare-earth ion with orbital moment $\mathbf{L}$
embedded in a free electron gas.}
\label{Fig0}
\end{figure}

\section{Local moments in a metallic host}

Rare-earth atoms generally appear in materials as triply charged cations.
Their partially filled 4f subshell governs their magnetic properties. The 4f
electrons only weakly interact with their environment~\cite%
{Liu1961,BookJensen} due to their small orbital radius and shielding by the
more extended and fully occupied $5s$ and $5p$ orbitals. This does not
exclude a significant exchange interaction: the conduction electrons of Pt
contacts activate the Gd moments in gadolinium gallium garnet (GGG)~\cite%
{OyanagiGGG}. The exchange interaction between a local spin with conduction
electrons of a metal host generates RKKY spin-density oscillations. Triply
charged Lanthanide anions have electronic configuration [Xe] 4f$^{n}$, where
the number of 4f electrons $n$ goes from $n=0$ for La$^{+3}$ to $n=14$ for Lu%
$^{+3}$. Except for $n=0$ (La$^{+3}$), $n=7$ (Gd$^{+3}$), and $n=14$ (Lu$%
^{+3}$), the intra-atomic spin-orbit interaction critically affects the
magnetic properties. Here we address a spin-orbit proximity effect of such a
magnetic moment embedded in a Fermi sea.

A partially occupied 4f subshell is characterized by a spin $\mathbf{S}$, an
orbital moment $\mathbf{L}$, and a total angular moment $\mathbf{J}=\mathbf{L%
}+\mathbf{S}$~\cite{Sievers,BookJensen}. For the basis $|\Psi \rangle\equiv|{%
S,L,J,J_{z}}\rangle$, $\mathbf{S}^{2}|\Psi\rangle=\hbar
^{2}S(S+1)|\Psi\rangle$, $\mathbf{L}^{2}|\Psi\rangle=\hbar^{2}L(L+1)|\Psi
\rangle$, $\mathbf{J}^{2}|\Psi\rangle=\hbar^{2}J(J+1)|\Psi\rangle$, $\hat{%
J_{z}}|\Psi\rangle=\hbar J_{z}|\Psi\rangle$, where $\hbar$ is the reduced
Planck constant. Hund's rules specify the quantum numbers $S$, $L$, and $J$
of the ground state manifold, while $J_{z}$ depends on the applied magnetic
and electric fields. Within a manifold of constant $S$, $L$, and $J$, the
Wigner-Eckart theorem ensures collinearity of all angular moment vectors: $%
\mathbf{S}=(g_{J}-1)\mathbf{J}$, $\mathbf{L}=(2-g_{J})\mathbf{J}$, and $%
\mathbf{L}+2\mathbf{S}=g_{J}\mathbf{J}$, where $%
g_{J}=3/2+[S(S+1)-L(L+1)]/[2J(J+1)]$ is the Land\'{e} g-factor.

We model the system as a single RE local moment embedded into a free
electron gas, which is appropriate for most dilute alloys. Conduction
electrons interact with the rare-earth spin and orbital moment via the Kondo
Hamiltonian~\cite{Kondo}. Here, we address equilibrium properties that are
affected by the spin-independent skew scattering but disregard external
current-induced phenomena such as the spin-Hall effect. We
operate in a regime above the Kondo temperature, and treat $\mathbf{J,S}$
and $\mathbf{L}$ as classical vectors. The strongly localized 4f orbital
radius governs the spatial extent of the coupling. When the 4f orbital
radius is much smaller than the typical wavelength of the conduction
electrons, the moment couples to free electrons by a contact interaction.

The $s$-$f$ exchange interaction in the Kondo Hamiltonian is similar to the $%
s$-$d$ Hamiltonian for $3d$-transition-metals~\cite%
{Liu1961,SFInteraction1,SFInteraction2,Kondo}. It reads 
\begin{equation}
H_{sf}=-\frac{J_{ex}}{\hbar^{2}}\delta^{\mathrm{4f}}\left( \mathbf{r}\right) 
\mathbf{S}\cdot\frac{\hbar\boldsymbol{\sigma}}{2},  \label{HamiltonianSF}
\end{equation}
where $\boldsymbol{\sigma}$ is the vector of Pauli matrices and $\delta ^{%
\mathrm{4f}}\left( \mathbf{r}\right) $ is a Dirac delta representing the
localized 4f subshell. In a free electron gas with Fermi wave number $k_{F}$%
, the exchange constant~\cite{Kondo} is $J_{ex}=2e^{2}A_{3}(0)/(7\epsilon
_{0}k_{F}^{2})$, where the radial integral~\cite{Kondo} $A_{h}(n)=\int_{0}^{%
\infty}dx_{1}x_{1}^{2}\int_{0}^{%
\infty}dx_{2}x_{2}^{2}j_{n}(x_{1})j_{n}(x_{2})x_{<}^{h}/x_{>}^{h+1}R(r_{1})R(r_{2}) 
$, with $x_{1}=k_{F}r_{1}$, $x_{2}=k_{F}r_{2}$, $x_{<}=\min(x_{1},x_{2})$,
and $x_{>}=\max(x_{1},x_{2})$. $A_{h}(n)$ can be evaluated numerically using
a Slater-type orbital for the radial part of the 4f wave function $R(r)\sim
r^{3}e^{-r/a}$ normalized over a large volume, $\int_{0}^{%
\infty}drr^{2}R^{2}(r)=1$. The constant $a$ is related to the 4f radius by $%
\langle r\rangle=\int_{0}^{\infty}drr^{3}R^{2}(r)=9a/2$. With $k_{F}=1.75$ 
\AA $^{-1}$ for Al and $\langle r\rangle=0.6$ \AA ~\cite{BookJensen}, $%
A_{3}(0)=0.33$ and $J_{ex}=5.6$ eV\thinspace\AA $^{3}$.

The so-called spin-\emph{in}dependent skew scattering affects the
trajectories of the electron charge and is responsible for the anomalous
Hall effect in metals with RE impurities~\cite{Fert,Kondo,Giovannini}. As
shown below, it also affects the ground state. Its Hamiltonian reads~\cite%
{Kondo} 
\begin{equation}
H_{\mathrm{skew}}=\mathbf{L}\cdot\left( \left[ \boldsymbol{\nabla}\eta\left( 
\mathbf{r}\right) \right] \times\frac{1}{i}\boldsymbol{\nabla }\right) 
\mathbb{I}_{2\times2},  \label{SpinIndpendenHam}
\end{equation}
where $\mathbb{I}_{2\times2}$ is the identity matrix in Pauli spin space,
and $\eta\left( \mathbf{r}\right) =\eta_{0}\delta^{\mathrm{4f}}\left( 
\mathbf{r}\right) $ with $\eta_{0}=9e^{2}\left[ A_{2}(1)-(5/9)A_{4}(1)\right]
\left( 140\epsilon_{0}\hbar k_{F}^{4}\right) ^{-1}$. For the parameters
introduced above, $A_{2}(1)\sim0.0885$, $A_{4}(1)=0.056$, and $%
\hbar\eta_{0}k_{F}^{2}=0.21$ eV \AA $^{3}$. Both exchange and
skew-scattering interactions are active in a volume $V_{\mathrm{4f}%
}\lesssim10\,\mathrm{\mathring{A}}^{3}$. The energy scales $\langle H_{%
\mathrm{skew}}\rangle\sim\hbar\eta_{0}k_{F}^{2}/V_{\mathrm{4f}}=\mathcal{O}%
\left( 10\,\mathrm{meV}\right) $ and $\langle H_{sf}\rangle\sim J_{ex}/V_{%
\mathrm{4f}}=\mathcal{O}\left( 100\,\mathrm{meV}\right) $ are consistent
with published values extracted from experiments, such as the Knight shift~%
\cite{StrengthSF}, electron spin resonance~\cite{RettoriStrengthSF}, and
magnetoresistance~\cite%
{FertStrengthSF,FertAndLevyStrengthSF,GiovaniStrengthSF}. $H_{\mathrm{skew}}$
deflects free electrons via an effective local force caused by the 4f
subshell with orbital angular momentum $\mathbf{L}$. Equation~(\ref%
{SpinIndpendenHam}) does not contain an explicit SOI parameter because we
operate in the limit of large 4f spin-orbit interaction that generates a
finite $\left\vert \mathbf{L}\right\vert $.

The 4f RE impurities in noble metals hybridize with 5d virtual-bound states
of the conduction electrons~\cite{GeneralModelForImpurities}, which can be
parameterized in terms of phase shifts of angular momentum scattering
channels~\cite{FertStrengthSF}. The enhancement of the
magnetic moments of pure RE metals~\cite%
{ExcessInPureMetals1,ExcessInPureMetals2,ExcessInPureMetals3} and in
RE-doped Ag and Au~\cite{Devine,Bak} has been attributed to those 5d virtual
bound states. Here we focus on the spin and orbital polarization induced by
the Kondo Hamiltonian on conduction electrons that we describe by plane
waves without truncating an expansion into spherical harmonics. To leading
order in the contact interaction, we may discard hybridization and
orthogonalization corrections.

Next, we discuss the RKKY spin polarization due to the $H_{sf}$ and the
response induced by $H_{\mathrm{skew}}$. We focus on ions with partially
filled 4f shells. Gd$^{3+}$ ($L=0$) can create an RKKY spin polarization,
but its $H_{\mathrm{skew}}$ vanishes. We do not address Eu$^{3+}$ since its
spin and orbital moment cancel in its ground ($J=0$) but not in excited
states.


\section{RKKY spin-density oscillations}

In the mean-field, local-density approximation, Eq.~(\ref{HamiltonianSF})
for an RE moment at the origin $\mathbf{r}=0$ becomes 
\begin{equation}
H_{sf}=-\frac{J_{ex}}{\hbar^{2}}\mathbf{s}(\mathbf{r}=0)\cdot\mathbf{S},
\label{Hsf}
\end{equation}
where $\mathbf{s}(\mathbf{r})\equiv\left\langle \Psi_{c}^{\dagger}(\mathbf{r}%
)\left\vert \hbar\boldsymbol{\sigma}/2\right\vert \Psi _{c}(\mathbf{r}%
)\right\rangle $ is the spin density of the conduction-electron wave
function $\Psi_{c}$. For a static moment and to leading order in $J_{ex}$,
we recover the RKKY spin density oscillations 
\begin{align}
\langle\mathbf{s}\rangle(r) & =\frac{J_{ex}}{\hbar^{2}}\chi(r)\mathbf{S}, \\
\chi(r) & =\frac{D_{e}\hbar^{2}}{16\pi r^{3}}\left[ \frac{\sin\left(
2k_{F}r\right) }{2k_{F}r}-\cos\left( 2k_{F}r\right) \right] ,
\label{EqLinearResponseSc}
\end{align}
where $\chi(r)$ is the susceptibility and $D_{e}=m_{e}k_{F}(\pi\hbar)^{-2} $
the density of states of the host metal at the Fermi energy. Figure~\ref%
{Fig1}(a) illustrates the characteristic RKKY oscillations in $r\chi(r)$
that contribute to the total spin magnetic moment $\mathbf{m}_{S}$ 
\begin{equation}
\mathbf{m}_{S}=-\gamma_{0}g_{S}\int d^{3}r\left[ \mathbf{S}\delta (\mathbf{r}%
)+\langle\mathbf{s}\rangle(\mathbf{r})\right] =-\gamma_{0}g_{S}\left(
1+G_{i}^{S}\right) \mathbf{S},
\end{equation}
where the bare $g$-factor is $g_{S}=2$, $\gamma_{0}=e/(2m_{e})$ is the
modulus of the gyromagnetic ratio, $-e$ is the electron charge, and the
constant $G_{i}^{S}=J_{ex}D_{e}/4$. For example, in Al, $G_{i}^{S}%
\approx0.13.$ The polarization cloud enhances the total spin magnetic moment
and $g$-factor by $G_{i}^{S}$, which can be observed via the imaginary part
of the spin-mixing conductance (effective field) at ferromagnet$|$normal
metal interfaces~\cite{crystalfield} or spin-dependent interfacial phase
shifts at ferromagnet$|$ superconductor interfaces~\cite{Cottet}.


\section{Rotational currents}

We now show that the Kondo Hamiltonian generates equilibrium charge-current
vortices around the impurity. In Fourier representation with linear momentum 
$\hbar\mathbf{q}$ and unperturbed wave function $\langle\mathbf{r}|\mathbf{q}%
\rangle=e^{i\mathbf{r}\cdot\mathbf{q}}/\sqrt{\Omega}$, the matrix elements
of the skew-type interaction read 
\begin{equation}
\left\langle \mathbf{q}+\mathbf{k}\right\vert H_{\mathrm{skew}}|\mathbf{q}%
\rangle=i\eta_{0}\Omega^{-1}e^{-2k^{2}a^{2}}(\mathbf{k}\times\mathbf{q})\cdot%
\mathbf{L},
\end{equation}
where $e^{-2k^{2}a^{2}}$ cuts off an ultraviolet divergence of a
delta-function potential. To leading order in $\eta_{0}$, we find a
spin-independent velocity field of conduction electrons~(see Appendix~\ref%
{AppendixVelocity}): 
\begin{equation}
\left\langle \mathbf{v}\left( \mathbf{r}\right) \right\rangle =\frac {%
\eta_{0}}{2\pi^{3}\hbar}\frac{F(r)}{r}\mathbf{L}\times\mathbf{\hat{r}},
\label{EqCurrentR}
\end{equation}
where 
\begin{align}
F(r) & =\frac{1}{ar\sqrt{2\pi}}\int_{0}^{\infty}\frac{dr^{\prime}}{r^{\prime
}}e^{-\frac{r^{\prime2}+r^{2}}{8a^{2}}}f_{1}(r,r^{\prime})f(r^{\prime}), \\
f_{1}(r,r^{\prime}) & =r^{\prime}r\cosh\left( \frac{r^{\prime}r}{4a^{2}}%
\right) -4a^{2}\sinh\left( \frac{r^{\prime}r}{4a^{2}}\right) , \\
f(x/k_{F}) & =\frac{2x\left( -9+2x^{2}\right) \cos\left( 2x\right) +\left(
9-14x^{2}\right) \sin\left( 2x\right) }{8(x/k_{F})^{6}},
\end{align}
and $x=k_{F}r$. In the delta-function $a\rightarrow0,$ the response function 
$f$ should used in Equation (\ref{EqCurrentR}) instead of $F$. Outside the
4f subshell, $F$ and $f$ are similar, see Fig.~\ref{Fig1}(b). $F$, $f$, and $%
\chi$ oscillate with wavenumber $2k_{F}$ and are approximated by $\cos\left(
2k_{F}r\right) /r^{3}$ far from the atom. The integrated modulus of the
velocity, 
\begin{equation}
\bar{v}_{s}=\int\left\vert \left\langle \mathbf{v}\right\rangle \right\vert
d^{3}r=0.01v_{F}\left( \frac{\left( 2-g_{J}\right) \sqrt{J\left( J+1\right) }%
}{4}\right) ,
\end{equation}
is of the order of a percent of the Fermi velocity $v_{F}=\hbar
k_{F}/m_{e}\sim2\times10^{6}$~m/s and $\bar{v}_{s}\sim10$~km/s for Al.

The radial density of the orbital angular momentum%
\begin{equation}
\langle \mathbf{l}\rangle (r)\equiv \int \int \frac{d\phi d\theta \sin
\theta }{4\pi }m_{e}\mathbf{r}\times \langle \mathbf{v}_{s}(\mathbf{r}%
)\rangle =\frac{m_{e}\eta _{0}F(r)}{3\pi ^{3}\hbar }\mathbf{L}.
\end{equation}%
\begin{figure}[t]
\includegraphics[width=8cm]{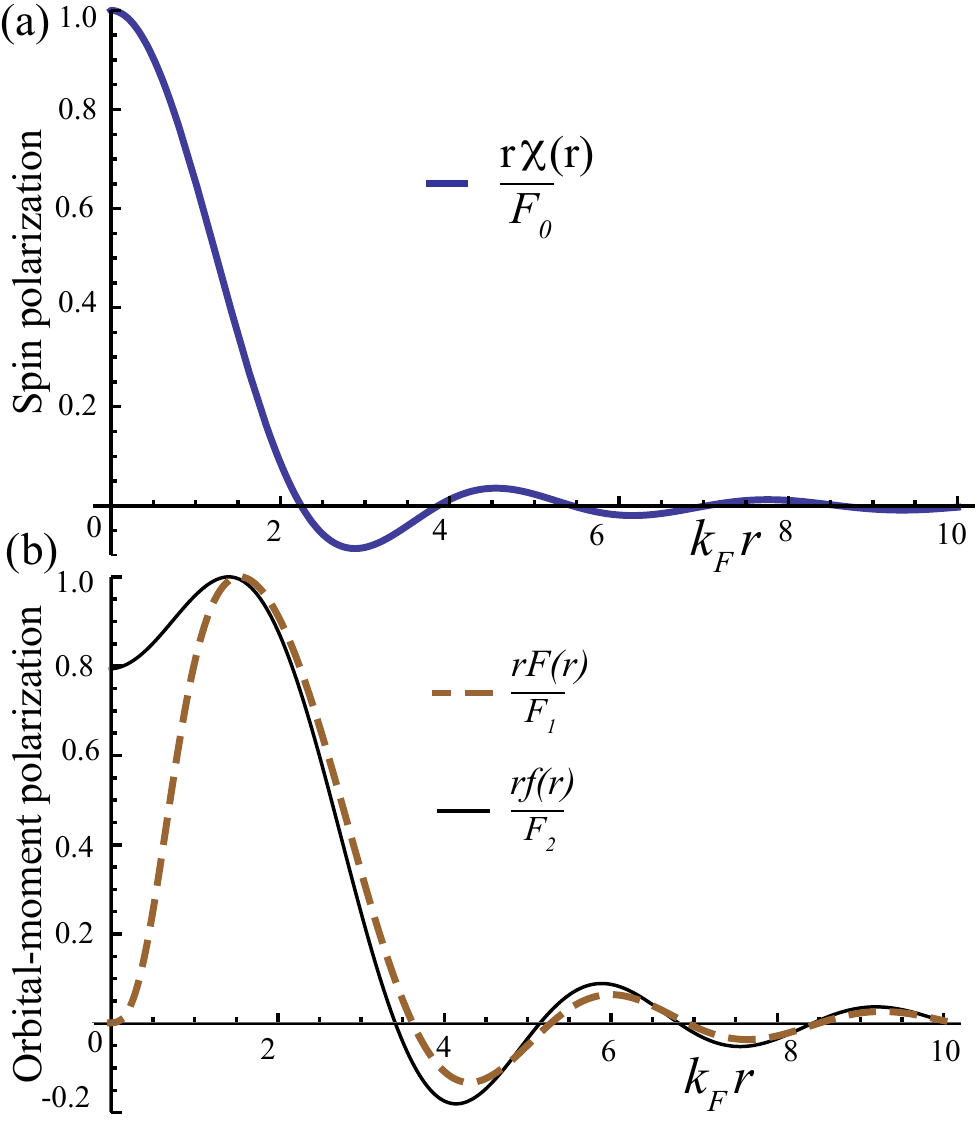}
\caption{Distribution of the spin and orbital polarizations close to a RE
local moment in the free electron gas. (a) RKKY spin density oscillations $r%
\protect\chi (r)/F_{0}$ normalized by $F_{0}=\lim_{r\rightarrow 0}r\protect%
\chi (r)$. (b) Distribution of the induced orbital moment obtained with the
regularized and non-regularized response functions, $rF(r)/F_{1}$ and $%
rf(r)/F_{2}$, respectively, normalized by their maximum values obtained
approximately for $k_{F}r=3/2$, $F_{1}=\max \left[ rF(r)\right] $ and $%
F_{2}=\max \left[ rf(r)\right] $. The response is dominantly paramagnetic
but oscillates with diamagnetic contributions. Far from the atom, the spin
and orbital responses share an oscillating algebraic decay $\cos \left(
2k_{F}r\right) /r^{3}.$ As this figure illustrates, the distributions have a
phase difference for small radius.}
\label{Fig1}
\end{figure}
is at equilibrium always collinear with $\mathbf{L}$ and parallel to it near
the origin. Both the orbital density of the rotational current and the RKKY
spin polarization decay algebraically and oscillate with increasing distance
from the origin [see Fig.~\ref{Fig1}(b)]. Therefore, the current response
has paramagnetic as well as diamagnetic contributions. Note that the spin
and orbital radial distributions have a phase difference for small radius,
as shown in Fig.~\ref{Fig1}.

The orbital magnetic moment is dressed by the electron gas%
\begin{equation}
\mathbf{m}_{L}=-\gamma _{0}g_{L}\int d^{3}r\left[ \mathbf{L}\delta (\mathbf{r%
})+\langle \mathbf{l}\rangle (\mathbf{r})\right] =-\gamma _{0}g_{L}\left(
1+G_{i}^{L}\right) \mathbf{L},
\end{equation}%
where the bare orbital $g$-factor $g_{L}=1$, and $G_{i}^{L}=2m_{e}\eta
_{0}k_{F}^{3}/\left( 3\pi ^{2}\hbar \right) $, or 
\begin{equation}
G_{i}^{L}=n_{\mathrm{4f}}\frac{E_{\eta }}{E_{F}},
\end{equation}%
where $n_{\mathrm{4f}}=V_{\mathrm{4f}}n_{0}$ is the number of
conduction electrons in the volume of the 4f subshell, $V_{\mathrm{4f}}\sim
10$~\r{A}$^{3}$, with $n_{0}=k_{F}^{3}/\left( 3\pi ^{2}\right) $ the metal
density. The energy of orbital-orbital coupling is $E_{\eta }=\hbar \eta
_{0}k_{F}^{2}/V_{\mathrm{4f}}$, while the host Fermi energy is $%
E_{F}=\hbar ^{2}k_{F}^{2}/\left( 2m_{e}\right) $. Thus, the strength of the
rotational current momentum, with respect to $\mathbf{L}$, is proportional
to the ratio of the orbital coupling and Fermi energies. The proportionality
constant is the average number of conduction electrons subject to the
coupling potential. For the present parameters we find $n_{\mathrm{4f}%
}\approx 1.8$, $E_{\eta }\approx 0.02$~eV, $E_{F}=11.7$~eV, and a relatively
small value $G_{i}^{L}\approx 1/300$.

The current vortex induces an \O rsted field 
\begin{equation}
\mathbf{B}_{\text{\textrm{\O }}}(\mathbf{r})=-\frac{e\mu _{0}}{4\pi }\int
d^{3}r^{\prime }\left\langle \mathbf{v}(\mathbf{r^{\prime }})\right\rangle
\times \frac{\mathbf{r}-\mathbf{r^{\prime }}}{|\mathbf{r}-\mathbf{r^{\prime }%
}|^{3}},  \label{EqMagnField}
\end{equation}%
where $\mu _{0}=4\pi \times 10^{-7}$\thinspace $\mathrm{J}/\left( \mathrm{mA}%
^{2}\right) $ is the magnetic permeability of free space. The
velocity and magnetic fields are proportional to the RE orbital momentum, $%
|\left\langle \mathbf{v}\right\rangle |\propto |\mathbf{B}_{\mathrm{\O }%
}|\propto \eta _{0}|\mathbf{L}|$. Far from the RE ion, $R\gg \langle
r\rangle $, 
\begin{equation}
\mathbf{B}_{\mathrm{\O }}(\mathbf{R})=\frac{\mu _{0}}{4\pi }\frac{3\left( 
\mathbf{m_{\mathrm{rc}}}\cdot \mathbf{R}\right) \mathbf{R}-R^{2}\mathbf{m_{%
\mathrm{rc}}}}{R^{5}}.
\end{equation}%
is the field generated by the magnetic dipole $\mathbf{m_{\mathrm{rc}}}%
=-\gamma _{0}G_{i}^{L}\mathbf{L}$. $\mathbf{B}_{\mathrm{\O }}$
is proportional to the dipolar magnetic field by the 4f orbital momentum ($%
\mathbf{B}_{\mathbf{L}}$), i.e., $\mathbf{B}_{\mathrm{\O }}=G_{i}^{L}\mathbf{%
B}_{\mathbf{L}}$, with $G_{i}^{L}\ll 1$. 

At the origin and for $k_{F}=1.75$/\r{A}%
\begin{align*}
\mathbf{B}_{\text{\textrm{\O }}}(\mathbf{0})& =-3.5\times 10^{-7}e\eta
_{0}\mu _{0}a^{-6}\left( \mathbf{L}/\hbar \right) , \\
& =0.06\text{ }\mathrm{T}\left( \frac{\left( 2-g_{J}\right) \sqrt{J\left(
J+1\right) }}{4}\right) .
\end{align*}%
\begin{figure}[b]
\includegraphics[width=8cm]{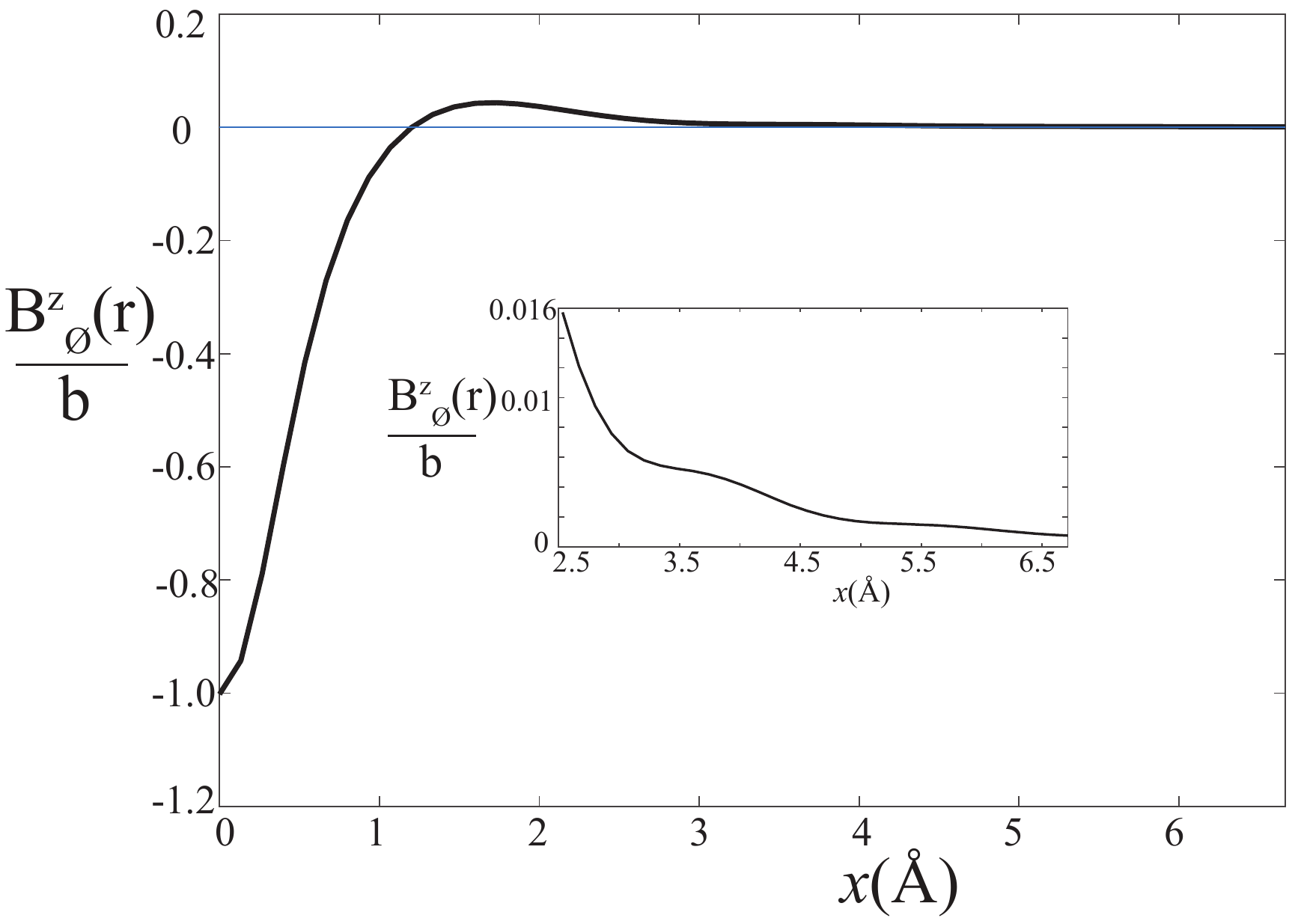}
\caption{The $z$-component of the \O rsted field near a RE atom in the free
electron gas. $B_{\text{\textrm{\O }}}^{z}=\mathbf{e_{z}}\cdot \mathbf{B}_{%
\text{\textrm{\O }}}$, as a function distance ($\mathbf{r}=x\mathbf{e_{x}}$%
), where $\mathbf{e_{j}}$ is the unit vector along the Cartesian axis $j$.
The field is normalized by its modulus at the origin $b=|\mathbf{B}_{\text{%
\textrm{\O }}}(\mathbf{0})|$.}
\label{FigField}
\end{figure}
Figure~\ref{FigField} shows the $z$-component of this field as a function of
distance $\mathbf{r}=x\mathbf{e_{x}}$, where the unit vectors $\mathbf{e_{x}}
$ and $\mathbf{e_{z}}$ pointing along the $x$- and $z$-Cartesian axis,
respectively. The field is negative close to the origin but turns positive
and decays to zero in an oscillatory fashion.

The field at the origin, $\mathbf{B}_{\text{\textrm{\O }}}(\mathbf{0})$,
couples to the local nuclear spin by the Zeeman interaction, that shifts the
nuclear magnetic resonance (NMR) frequency by $\Delta \omega =\gamma
_{N}\left\vert \mathbf{B}_{\text{\textrm{\O }}}(\mathbf{0})\right\vert \sim 5
$ MHz, where $\gamma _{N}$ is the nuclear gyromagnetic ratio. In an NMR
experiments, a constant magnetic field $B_{0}\mathbf{e_{z}}$ polarizes the
nuclear moments as well as the 4f moments along the $z$-direction. The
Knight shift $K$ produced by the current and parameterized by the ratio of
the internal and applied magnetic fields~\cite{SolidBookKittel}, at low
temperatures ($T\lesssim 1$ K) is 
\begin{equation}
K\equiv \frac{|\mathbf{B}_{\text{\textrm{\O }}}(0)|}{B_{0}}=-0.3\%\left( 
\frac{\left( 2-g_{J}\right) \sqrt{J\left( J+1\right) }}{4}\right) \left( 
\frac{20\text{\thinspace \textrm{T}}}{B_{0}}\right) .  \label{KnightShift1}
\end{equation}%
where we assume full polarization of the magnetic moment of the 4f subshell.
The nuclear magnetic resonance frequency is typically in the 100 MHz regime
for applied (constant) fields of $B_{0}\sim 10$~T. For a given rf frequency
we therefore predict different resonance magnetic fields for rare earth
impurities in an insulating and metallic host. We hope that our results
stimulate experiments that can identify the ground state current vortices.

Orbital contributions to the Knight shift have been predicted before~\cite%
{Devine,KnightShiftInSystemsWithSOC}, holding virtual bound states of
conduction electrons at RE impurities in a metal host responsible~\cite%
{Devine}. These theories are not compatible with our model since they
predict effects for half-filled shells without orbital moment (Gd). 

\section{Conclusions}

We predict that RE local moments interact with the conduction electrons of a
metallic host to generate both an oscillating spin density and charge
current, the latter by the spin-\emph{in}dependent skew-scattering
interaction. The radial distribution of the induced velocity field
oscillates with the same period as the RKKY spin polarization induced by the
local exchange interaction. A finite 4f orbital moment $\mathbf{L}$ is
necessary to form the current in the electron gas. Therefore, the predicted
magnetic field and Knight shift depend linearly with $\left\vert \mathbf{L}%
\right\vert \propto\left( 2-g_{J}\right) \left\vert \mathbf{J}\right\vert $
and vanish for RE-doped insulators. Furthermore, the induced magnetic field
depends on the atomic number via the Land\'{e} g-factor ground-state quantum
numbers. Several approximations are crude, but we are confident about the
predicted trends. Relativistic first-principles calculations of open 4f
subshells in a metal host should improve the accuracy of the predictions.

The \O rsted fields generated by RE impurities at the surface of a metal
with sufficiently large Fermi wavelength can be measured in
principle by scanning magnetometries based on nano-scale superconducting
quantum interference devices (nanoSQUID)~\cite{NanoSquid2} or optically
read-out nitrogen-vacancy (NV)-centers~\cite{NVCenters1,NVCenters2}. RE
impurities adsorbed at a surface 2D electron gas, or graphene monolayer are
promising candidate systems to image the predicted equilibrium current
vortices.

\section*{Acknowledgments}

We thank Jiang Xiao and Koichi Oyanagi for fruitful discussions. This
research was supported by JSPS KAKENHI Grant No. 19H006450, Postdoctorado
FONDECYT 2019 Folio 3190030, and PUTIQ1 Grant of Universitas Indonesia
NKB-1369/UN2.RST/HKP.05.00/2020.

\appendix

\section{Rotational currents}

\label{AppendixVelocity} Here we derive the linear response of a simple
metal to a rare-earth (RE) magnetic impurity characterized by the classical
vectors $\mathbf{S}$, $\mathbf{L}$, and $\mathbf{J}$, i.e. the spin,
orbital, and total angular momenta, respectively. The conduction-electron
orbital angular momentum density $\mathbf{l}=m_{e}\mathbf{r}\times \mathbf{v}%
\left( \mathbf{r}\right) \boldsymbol{\ }$relative to the local moment at the
origin, where $m_{e}\mathbf{v\left( \mathbf{r}\right) }$ is the linear
momentum density, reads in second quantization (see also Appendix~\ref%
{Sec:SI:orbital}) 
\begin{equation}
\mathbf{l}(\mathbf{r})=\hbar \mathbf{r}\times \frac{1}{\Omega }\sum_{\mathbf{%
pq}\gamma }e^{i\left( \mathbf{q}-\mathbf{p}\right) \cdot \mathbf{r}}\frac{%
\mathbf{p}+\mathbf{q}}{2}a_{\mathbf{p}\gamma }^{\dagger }a_{\mathbf{q}\gamma
}.
\end{equation}%
We define the expectation values $\left\langle \mathbf{v}\right\rangle =%
\mathrm{Tr}\left[ {\rho }\mathbf{v}\right] $, where Tr stands for the trace,
and $\rho $ is the density matrix of the full Hamiltonian. With
time-evolution operator in the interaction picture, ${U}(t)\equiv \exp \left[
-(i/\hbar )\int_{t_{0}}^{t}H_{\mathrm{skew}}(t^{\prime })dt^{\prime }\right] 
$, the total density matrix ${\rho }$ can be written in terms of the ground
state density matrix ${\rho }_{0}$ of the unperturbed free electron gas with
Hamiltonian $H_{0}$ and the regularized skew scattering Hamiltonian 
\begin{equation}
H_{\mathrm{skew}}=\frac{i\eta _{0}}{\Omega }\sum_{\mathbf{k}\mathbf{q}%
^{\prime }\gamma ^{\prime }}e^{-2k^{2}a^{2}}a_{\mathbf{q}^{\prime }+\mathbf{k%
}\ \gamma ^{\prime }}^{\dagger }a_{\mathbf{q}^{\prime }\gamma ^{\prime }}(%
\mathbf{k}\times \mathbf{q}^{\prime })\cdot \mathbf{L},
\end{equation}%
where the constant $a=2\langle r\rangle /9$, related to the 4f subshell
radius $\langle r\rangle \sim 0.6$\AA , accounts for the finite spatial
extension of the 4f Slater-type orbital $R(r)\propto r^{3}e^{-r/a}$. The
exponential $e^{-2k^{2}a^{2}}$ cuts off an ultraviolet divergence that would
arrive for a delta-function perturbation. For an Al host metal, $\hbar \eta
_{0}k_{F}^{2}=0.21$ eV \AA $^{3}$. Then,%
\begin{align}
\left\langle \mathbf{v}(\mathbf{r})\right\rangle =\mathrm{Tr}& \left[ {\rho }%
_{0}\hat{U}^{-1}(t)\mathbf{v}\hat{U}(t)\right] ,  \notag \\
\approx \mathrm{Tr}& \left[ {\rho }_{0}\left( 1+\frac{i}{\hbar }%
\int_{-\infty }^{t}H_{\mathrm{skew}}(t^{\prime })dt^{\prime }\right) \mathbf{%
v}\right.  \notag \\
& \left. \cdot \left( 1-\frac{i}{\hbar }\int_{-\infty }^{t}H_{\mathrm{skew}%
}(t^{\prime })dt^{\prime }\right) \right] ,  \notag \\
=\frac{i}{\hbar }& \left\langle \int_{-\infty }^{t}dt^{\prime }\left[ H_{%
\mathrm{skew}}(t^{\prime }),\mathbf{v}(\mathbf{r})\right] \right\rangle _{0},
\end{align}%
where $\langle {A}\rangle _{0}=\mathrm{Tr}\left[ {\rho }_{0}{A}\right] $.
This leads to%
\begin{align}
\left\langle \mathbf{v}(\mathbf{r})\right\rangle =& -\frac{i}{m_{e}\Omega }%
\int_{-\infty }^{t}dt^{\prime }\sum_{\mathbf{pq}\gamma }e^{i(\mathbf{q}-%
\mathbf{p})\cdot \mathbf{r}}\frac{\mathbf{p}+\mathbf{q}}{2}  \notag \\
& \left\langle \left[ a_{\mathbf{p}\gamma }^{\dagger }(t)a_{\mathbf{q}\gamma
}(t),H_{\mathrm{skew}}\ (t^{\prime })\right] \right\rangle _{0}  \notag \\
=& \frac{\eta _{0}}{m_{e}\Omega ^{2}}\sum_{\mathbf{k}\mathbf{q}^{\prime
}\gamma ^{\prime }}e^{-2k^{2}a^{2}}\sum_{\mathbf{pq}\gamma }e^{i(\mathbf{q}-%
\mathbf{p})\cdot \mathbf{r}}\frac{\mathbf{p}+\mathbf{q}}{2}(\mathbf{k}\times 
\mathbf{q}^{\prime })\cdot \mathbf{L}  \notag \\
& \int_{-\infty }^{t}dt^{\prime }\left\langle \left[ a_{\mathbf{p}\gamma
}^{\dagger }(t)a_{\mathbf{q}\gamma }(t),a_{\mathbf{q}^{\prime }+\mathbf{k}\
\gamma ^{\prime }}^{\dagger }(t^{\prime })a_{\mathbf{q}^{\prime }\gamma
^{\prime }}(t^{\prime })\right] \right\rangle _{0}.  \label{EqCurrenPrevXi}
\end{align}%
The susceptibility is 
\begin{equation}
\chi (t-t^{\prime })=\Theta (t-t^{\prime })\sum_{\gamma \gamma ^{\prime
}}\left\langle \left[ a_{\mathbf{p}\gamma }^{\dagger }(t)a_{\mathbf{q}\gamma
}(t),a_{\mathbf{q}^{\prime }+\mathbf{k}\gamma ^{\prime }}^{\dagger
}(t^{\prime })a_{\mathbf{q}^{\prime }\gamma ^{\prime }}(t^{\prime })\right]
\right\rangle _{0},
\end{equation}%
where $\Theta $ is the Heaviside step function with time derivative 
\begin{align}
& \partial _{t}\chi (t-t^{\prime })=\delta (t-t^{\prime })\sum_{\gamma
\gamma ^{\prime }}\left\langle \left[ a_{\mathbf{p}\gamma }^{\dagger }(t)a_{%
\mathbf{q}\gamma }(t),a_{\mathbf{q}^{\prime }+\mathbf{k}\gamma ^{\prime
}}^{\dagger }(t^{\prime })a_{\mathbf{q}^{\prime }\gamma ^{\prime
}}(t^{\prime })\right] \right\rangle _{0}  \notag \\
& +\Theta (t-t^{\prime })\sum_{\gamma \gamma ^{\prime }}\left\langle \left[
\partial _{t}\left( a_{\mathbf{p}\gamma }^{\dagger }(t)a_{\mathbf{q}\gamma
}(t)\right) ,a_{\mathbf{q}^{\prime }+\mathbf{k}\gamma ^{\prime }}^{\dagger
}(t^{\prime })a_{\mathbf{q}^{\prime }\gamma ^{\prime }}(t^{\prime })\right]
\right\rangle _{0}.
\end{align}%
$\partial _{t}\left[ a_{\mathbf{p}\gamma }^{\dagger }(t)a_{\mathbf{q}\gamma
}(t)\right] $ can be calculated by the Heisenberg equation for the electron
gas $H_{0}=\sum_{\mathbf{k}\gamma }\epsilon _{\mathbf{k}}a_{\mathbf{k}\gamma
}^{\dagger }a_{\mathbf{k}\gamma }$ with parabolic dispersion relation $%
\epsilon _{\mathbf{k}}=\hbar ^{2}\mathbf{k}^{2}/(2m_{e})$, 
\begin{align}
{\partial _{t}\left( a_{\mathbf{p}\gamma }^{\dagger }a_{\mathbf{q}\gamma
}\right) }& =\frac{1}{i\hbar }\left[ a_{\mathbf{p}\gamma }^{\dagger }a_{%
\mathbf{q}\gamma },H_{0}\right]  \notag \\
& =\frac{1}{i\hbar }\left[ a_{\mathbf{p}\gamma }^{\dagger }a_{\mathbf{q}%
\gamma },\sum_{\mathbf{k}\gamma ^{\prime }}\epsilon _{\mathbf{k}}a_{\mathbf{k%
}\gamma ^{\prime }}^{\dagger }a_{\mathbf{k}\gamma ^{\prime }}\right]  \notag
\\
& =-\frac{i}{\hbar }\left( \epsilon _{\mathbf{q}}-\epsilon _{\mathbf{p}%
}\right) a_{\mathbf{p}\gamma }^{\dagger }a_{\mathbf{q}\gamma },
\end{align}%
and 
\begin{align}
& \sum_{\gamma \gamma ^{\prime }}\left[ a_{\mathbf{p}\gamma }^{\dagger }a_{%
\mathbf{q}\gamma },a_{\mathbf{q}^{\prime }+\mathbf{k}\gamma ^{\prime
}}^{\dagger }a_{\mathbf{q}^{\prime }\gamma ^{\prime }}\right]  \notag \\
& =\sum_{\gamma }\left( \delta _{\mathbf{q},\mathbf{q}^{\prime }+\mathbf{k}%
}a_{\mathbf{p}\gamma }^{\dagger }a_{\mathbf{q}-\mathbf{k}\gamma }-\delta _{%
\mathbf{p},\mathbf{q}^{\prime }}a_{\mathbf{p}+\mathbf{k}\gamma }^{\dagger
}a_{\mathbf{q}\gamma }\right) .
\end{align}%
$\chi $ then satisfies the equation of motion 
\begin{align}
& \left( \partial _{t}+\frac{i}{\hbar }\left( \epsilon _{\mathbf{q}%
}-\epsilon _{\mathbf{p}}\right) \right) \boldsymbol{\chi }(t-t^{\prime }) 
\notag \\
& =\delta (t-t^{\prime })\left\langle \sum_{\gamma }\left( \delta _{\mathbf{q%
},\mathbf{q}^{\prime }+\mathbf{k}}a_{\mathbf{p}\gamma }^{\dagger }a_{\mathbf{%
q}-\mathbf{k}\gamma }-\delta _{\mathbf{p},\mathbf{q}^{\prime }}a_{\mathbf{p}+%
\mathbf{k}\gamma }^{\dagger }a_{\mathbf{q}\gamma }\right) \right\rangle _{0}.
\end{align}%
In the frequency domain, with $\chi (t)=(2\pi )^{-1}\int d\omega \chi
(\omega )e^{-i\omega t}$ 
\begin{align}
\chi (\omega )& =i\hbar \frac{\sum_{\gamma }\left\langle \delta _{\mathbf{q},%
\mathbf{q}^{\prime }+\mathbf{k}}a_{\mathbf{p}\gamma }^{\dagger }a_{\mathbf{q}%
-\mathbf{k}\gamma }-\delta _{\mathbf{p},\mathbf{q}^{\prime }}a_{\mathbf{p}+%
\mathbf{k}\gamma }^{\dagger }a_{\mathbf{q}\gamma }\right\rangle _{0}}{%
\epsilon _{\mathbf{p}}-\epsilon _{\mathbf{q}}+\hbar \omega +i0^{+}}  \notag
\\
& =2i\hbar \frac{\delta _{\mathbf{q},\mathbf{q}^{\prime }+\mathbf{k}}\delta
_{\mathbf{p},\mathbf{q}-\mathbf{k}}f_{\mathbf{p}}-\delta _{\mathbf{p},%
\mathbf{q}^{\prime }}\delta _{\mathbf{q},\mathbf{p}+\mathbf{k}}f_{\mathbf{q}}%
}{\epsilon _{\mathbf{p}}-\epsilon _{\mathbf{q}}+\hbar \omega +i0^{+}},
\end{align}%
where $f_{\mathbf{p}}$ is the (spin-degenerate) Fermi-Dirac distribution.
Substituting $\chi $ into Eq.~(\ref{EqCurrenPrevXi}) after transformation
into the frequency domain and in the steady state ($\omega \rightarrow 0$) 
\begin{align}
\left\langle \mathbf{v}\left( \mathbf{r}\right) \right\rangle =& \frac{\eta
_{0}}{m_{e}\Omega ^{2}}\sum_{\mathbf{k}\mathbf{q}^{\prime
}}e^{-2k^{2}a^{2}}\sum_{\mathbf{pq}}e^{i(\mathbf{q}-\mathbf{p})\cdot \mathbf{%
r}}\left( \mathbf{p}+\mathbf{q}\right)  \notag \\
& \cdot \left[ (\mathbf{k}\times \mathbf{q}^{\prime })\cdot \mathbf{L}\right]
i\hbar \frac{\delta _{\mathbf{q},\mathbf{q}^{\prime }+\mathbf{k}}\delta _{%
\mathbf{p},\mathbf{q}-\mathbf{k}}f_{\mathbf{p}}-\delta _{\mathbf{p},\mathbf{q%
}^{\prime }}\delta _{\mathbf{q},\mathbf{p}+\mathbf{k}}f_{\mathbf{q}}}{%
\epsilon _{\mathbf{p}}-\epsilon _{\mathbf{q}}+i0^{+}}  \notag \\
=& \frac{i\eta _{0}\hbar }{m_{e}\Omega ^{2}}\sum_{\mathbf{pq}}e^{-2a^{2}|%
\mathbf{p}-\mathbf{q}|^{2}}e^{i(\mathbf{q}-\mathbf{p})\cdot \mathbf{r}%
}\left( \mathbf{p}+\mathbf{q}\right)  \notag \\
& \left( \mathbf{L}\cdot \frac{-\mathbf{p}\times \mathbf{q}f_{\mathbf{p}}-%
\mathbf{q}\times \mathbf{p}f_{\mathbf{q}}}{\epsilon _{\mathbf{p}}-\epsilon _{%
\mathbf{q}}+i0^{+}}\right)  \notag \\
=& \frac{i\eta _{0}\hbar }{m_{e}}\int \frac{d^{3}{p}}{(2\pi )^{3}}\int \frac{%
d^{3}{q}}{(2\pi )^{3}}e^{-2a^{2}|\mathbf{p}-\mathbf{q}|^{2}}e^{i(\mathbf{q}-%
\mathbf{p})\cdot \mathbf{r}}  \notag \\
& \left( \mathbf{p}+\mathbf{q}\right) \left[ \mathbf{L}\cdot (\mathbf{q}%
\times \mathbf{p})\right] \frac{f_{\mathbf{p}}-f_{\mathbf{q}}}{\epsilon _{%
\mathbf{p}}-\epsilon _{\mathbf{q}}+i0^{+}}  \notag \\
=& \frac{i\eta _{0}\hbar }{m_{e}}\int \frac{d^{3}{p}}{(2\pi )^{3}}\int \frac{%
d^{3}{q}}{(2\pi )^{3}}e^{-2a^{2}|\mathbf{p}-\mathbf{q}|^{2}}e^{i(\mathbf{q}-%
\mathbf{p})\cdot \mathbf{r}}  \notag \\
& \left( \mathbf{p}+\mathbf{q}\right) \left[ \mathbf{L}\cdot (\mathbf{q}%
\times \mathbf{p})\right] \frac{f_{\mathbf{p}}}{\epsilon _{\mathbf{p}%
}-\epsilon _{\mathbf{q}}+i0^{+}}+\mathrm{c.c.},
\label{EqCaluclationCurledCurr1}
\end{align}%
where c.c. stands for the complex conjugate of the other terms.

Using 
\begin{equation}
e^{-2a^{2}\left\vert \mathbf{p}-\mathbf{q}\right\vert ^{2}}=\frac{\sqrt{2}}{%
32\pi ^{3/2}a^{3}}\int d^{3}{r^{\prime }}e^{i\left( \mathbf{q}-\mathbf{p}%
\right) \cdot \mathbf{r}^{\prime }}e^{-r^{\prime 2}/(8a^{2})},
\end{equation}%
we can write the regularized velocity, $\left\langle \mathbf{v}\left( 
\mathbf{r}\right) \right\rangle $, as the integral of the non-regularized
one, $\left\langle \mathbf{v}_{\infty }\left( \mathbf{r}\right)
\right\rangle $. The later has a divergence in the origin due to the delta
nature of the skew scattering when $a\rightarrow 0$, as shown later. 
\begin{align}
\left\langle \mathbf{v}\left( \mathbf{r}\right) \right\rangle =& \frac{\sqrt{%
2}}{32\pi ^{3/2}a^{3}}\int d^{3}{r^{\prime }}e^{-r^{\prime
2}/(8a^{2})}\left\langle \mathbf{v}_{\infty }\left( \mathbf{r}+\mathbf{r}%
^{\prime }\right) \right\rangle ,  \notag \\
\left\langle \mathbf{v}_{\infty }\left( \mathbf{r}\right) \right\rangle =&
i\eta _{0}\frac{\hbar }{m_{e}}\int \frac{d^{3}{p}}{(2\pi )^{3}}\int \frac{%
d^{3}{q}}{(2\pi )^{3}}e^{i(\mathbf{q}-\mathbf{p})\cdot \mathbf{r}}\left( 
\mathbf{p}+\mathbf{q}\right)  \notag \\
& \cdot \left[ \mathbf{L}\cdot (\mathbf{q}\times \mathbf{p})\right] \frac{f_{%
\mathbf{p}}}{\epsilon _{\mathbf{p}}-\epsilon _{\mathbf{q}}+i0^{+}}+\mathrm{%
c.c.},  \label{EqReg}
\end{align}

The angular part of the integral over $\mathbf{q}=q\left( \sin \theta _{%
\mathbf{q}}\cos \phi _{\mathbf{q}}\mathbf{\hat{x}}+\sin \theta _{\mathbf{q}%
}\sin \phi _{\mathbf{q}}\mathbf{\hat{y}}+\cos \theta _{\mathbf{q}}\mathbf{%
\hat{z}}\right) $ reads 
\begin{align}
I_{1}& \equiv i\int_{0}^{\pi }d\theta _{\mathbf{q}}\sin \theta _{\mathbf{q}%
}\int_{0}^{2\pi }d\phi _{\mathbf{q}}e^{i\mathbf{q}\cdot \mathbf{r}}\left( 
\mathbf{q}+\mathbf{p}\right) \left[ \mathbf{L}\cdot (\mathbf{q}\times 
\mathbf{p})\right]   \notag \\
& =\frac{4\pi i}{qr^{3}}\left[ qr\cos \left( qr\right) -\sin \left(
qr\right) \right] \left[ \mathbf{L}\times \mathbf{p}+i\left( \mathbf{r}\cdot 
\mathbf{J}\times \mathbf{p}\right) \mathbf{p}\right]   \notag \\
& -\frac{4\pi i}{qr^{3}}\left[ 3qr\cos \left( qr\right) -\left(
3-q^{2}r^{2}\right) \sin \left( qr\right) \right] \left( \mathbf{\hat{r}}%
\cdot \mathbf{L}\times \mathbf{p}\right) \mathbf{\hat{r}},
\end{align}%
such that the angular integral over $\mathbf{p}=p\left( \sin \theta _{%
\mathbf{p}}\cos \phi _{\mathbf{p}}\mathbf{\hat{x}}+\sin \theta _{\mathbf{p}%
}\sin \phi _{\mathbf{p}}\mathbf{\hat{y}}+\cos \theta _{\mathbf{p}}\mathbf{%
\hat{z}}\right) $ of the previous expression is 
\begin{align}
& \int_{0}^{\pi }d\theta _{\mathbf{p}}\sin \theta _{\mathbf{p}%
}\int_{0}^{2\pi }d\phi _{\mathbf{p}}e^{-i\mathbf{p}\cdot \mathbf{r}}I_{1} \\
& =-\frac{32\pi ^{2}}{pqr^{5}}\left[ pr\cos \left( pr\right) -\sin \left(
pr\right) \right] \left[ qr\cos \left( qr\right) -\sin \left( qr\right) %
\right] \mathbf{L}\times \mathbf{\hat{r}},  \notag
\end{align}%
which reveals the rotational (i.e., $\propto \mathbf{L}\times \mathbf{\hat{r}%
}$) character of the current. 
\begin{align}
\left\langle \mathbf{v}_{\infty }\right\rangle & =-\frac{\eta _{0}}{\pi
^{4}\hbar r^{5}}\mathbf{L}\times \mathbf{\hat{r}}\int_{0}^{k_{F}}dpp\left[
pr\cos \left( pr\right) -\sin \left( pr\right) \right]  \\
& \int_{0}^{\infty }dq\frac{q\left[ qr\cos \left( qr\right) -\sin \left(
qr\right) \right] }{p^{2}-q^{2}+i0^{+}}+\text{c.c.}
\end{align}%
Using $\cos \left( qr\right) =\left( e^{iqr}+e^{-iqr}\right) /2$ and $\sin
\left( qr\right) =\left( e^{iqr}-e^{-iqr}\right) /(2i)$ 
\begin{align}
& \int_{0}^{\infty }dq\frac{q\left[ qr\cos \left( qr\right) -\sin \left(
qr\right) \right] }{p^{2}-q^{2}+i0^{+}}  \notag \\
& =-\frac{1}{4}\int_{-\infty }^{\infty }dqq\frac{e^{iqr}\left( qr+i\right)
+e^{-iqr}\left( qr-i\right) }{q^{2}-\left( p+i0^{+}\right) ^{2}}
\end{align}%
the integral over $q$ can be carried out by a contour integral in the
complex plane. For $r>0$ only the poles with a positive (negative) imaginary
part contribute for integrands containing $e^{iqr}$ ($e^{-iqr}$), 
\begin{equation}
\int_{0}^{\infty }dq\frac{q\left[ qr\cos \left( qr\right) -\sin \left(
qr\right) \right] }{p^{2}-q^{2}+i0^{+}}=-\frac{\pi i}{2}\left( pr+i\right)
e^{ipr}.
\end{equation}%
and 
\begin{align}
\left\langle \mathbf{v}_{\infty }\right\rangle & =\frac{\eta _{0}}{2\pi
^{3}\hbar r^{5}}\mathbf{L}\times \mathbf{\hat{r}}\int_{0}^{k_{F}}dpp\left[
pr\cos \left( pr\right) -\sin \left( pr\right) \right] \left( ipr-1\right)
e^{ipr}+\text{c.c.}  \notag \\
& =-\frac{\eta _{0}}{\pi ^{3}\hbar r^{5}}\mathbf{L}\times \mathbf{\hat{r}}%
\int_{0}^{k_{F}}dpp\left[ pr\cos \left( pr\right) -\sin \left( pr\right) %
\right]   \notag \\
& \left[ \cos \left( pr\right) +pr\sin \left( pr\right) \right] 
\end{align}%
The $p$ integral is straightforward leading to 
\begin{align}
\left\langle \mathbf{v}_{\infty }(\mathbf{r})\right\rangle & =\frac{\eta _{0}%
}{2\pi ^{3}\hbar }\frac{f(r)}{r}\mathbf{L}\times \mathbf{\hat{r}},
\label{EqVelApp} \\
f(x/k_{F})& =\frac{2x\left( -9+2x^{2}\right) \cos \left( 2x\right) +\left(
9-14x^{2}\right) \sin \left( 2x\right) }{8(x/k_{F})^{6}},
\end{align}%
where $x=k_{F}r$. $f(r)$ oscillates with wavenumber $2k_{F}$ as expected for
the response of a degenerate electron gas. Moreover, for $r\gg \langle
r\rangle $, $f(r)\propto \cos \left( 2k_{F}r\right) /r^{3}$, as well known
from the RKKY spin polarization. Finally, the divergence at the origin $%
\lim_{r\rightarrow 0}\left\vert \left\langle \mathbf{v}_{\infty }(\mathbf{r}%
)\right\rangle \right\vert =\infty $ comes from a delta-function skew
scattering potentials ($a\rightarrow 0)$ and can be avoided by using Eq.~(%
\ref{EqReg}), 
\begin{align}
\left\langle \mathbf{v}\left( \mathbf{r}\right) \right\rangle & =\frac{\eta
_{0}}{2\pi ^{3}\hbar }\frac{F(r)}{r}\mathbf{L}\times \mathbf{\hat{r}}, \\
F(r)& =\frac{1}{ar\sqrt{2\pi }}\int_{0}^{\infty }\frac{dr^{\prime }}{%
r^{\prime }}e^{-\frac{r^{\prime 2}+r^{2}}{8a^{2}}}  \notag \\
& \cdot \left[ r^{\prime }r\cosh \left( \frac{r^{\prime }r}{4a^{2}}\right)
-4a^{2}\sinh \left( \frac{r^{\prime }r}{4a^{2}}\right) \right] f(r^{\prime
}),
\end{align}%
Both response functions are plotted in Fig.~\ref{FigA1}. 
\begin{figure}[t]
\includegraphics[width=9cm]{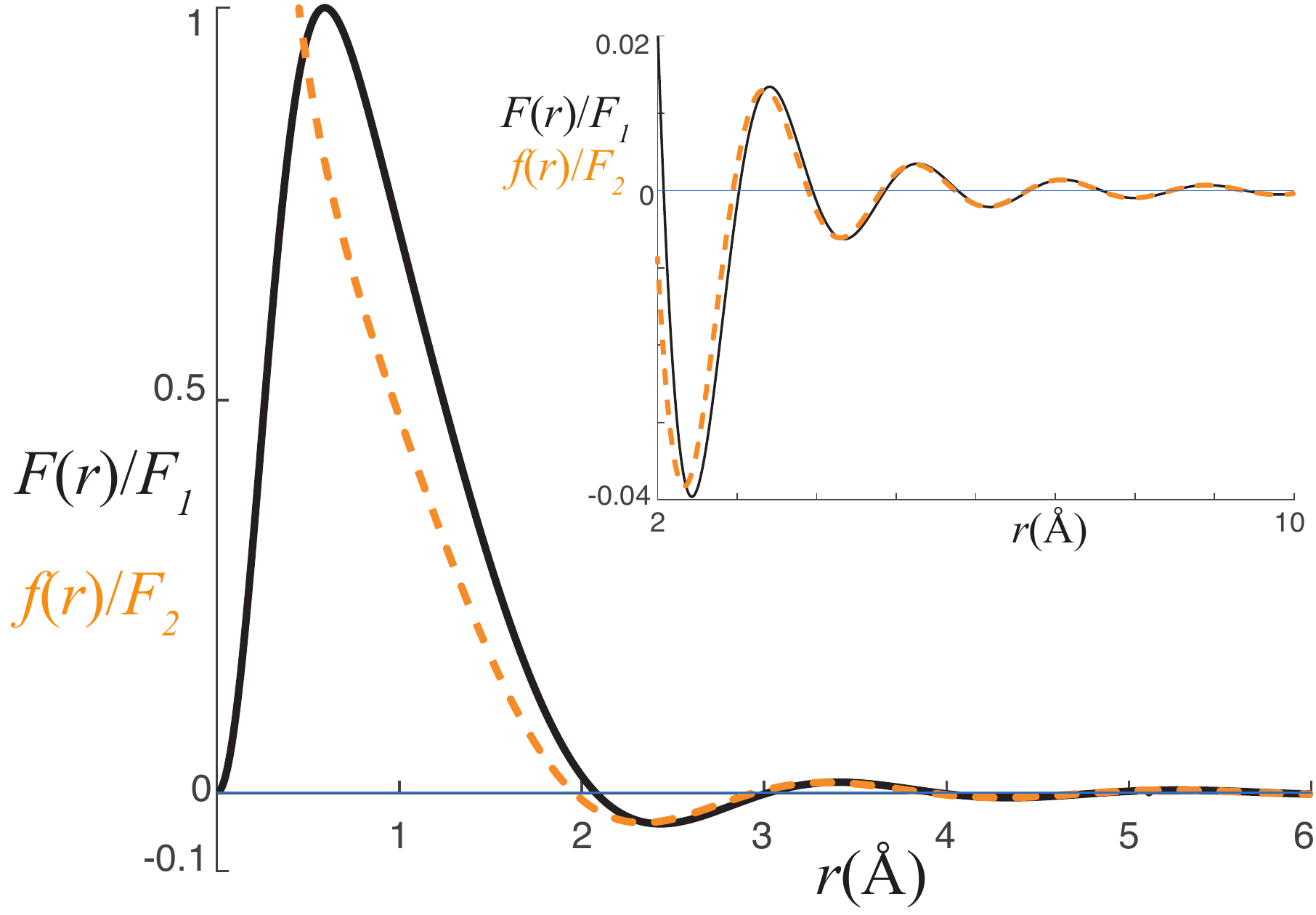}
\caption{Response functions $F(r)$ and $f(r)$ that describe the regularized
and non-regularized velocities, respectively. They are normalized by the
maximum value of $F_{1}\equiv $ max$[F(r)]$ and by $F_{2}=F_{1}f\left(
r_{0}\right) /F\left( r_{0}\right) $, such that they have the same value at
the (arbitrarily chosen) point $r_{0}=3.34$ \r{A}. The inset shows the
behavior of the curves for large distances. As this figure illustrates, the
curves have the same feature outside the rare-earth atom. However, only the
regularized function $f$ is finite inside the 4f subshell.}
\label{FigA1}
\end{figure}

The angular average of the orbital moment density 
\begin{equation}
\langle \mathbf{l}\rangle (r)\equiv \frac{1}{4\pi }\int_{0}^{2\pi }d\phi
\int_{0}^{\pi }d\theta \sin \theta m_{e}\mathbf{r}\times \left\langle 
\mathbf{v}\left( \mathbf{r}\right) \right\rangle =\frac{m_{e}\eta _{0}F(r)}{%
3\pi ^{3}\hbar }\mathbf{L}.
\end{equation}%
with integrated value 
\begin{equation*}
\int d^{3}r\langle \mathbf{l}\rangle (r)=\frac{4m_{e}\eta _{0}}{3\pi
^{2}\hbar }\int_{0}^{\infty }drr^{2}F(r)\mathbf{L}=G_{i}^{L}\mathbf{L},
\end{equation*}%
and 
\begin{equation}
\int_{0}^{\infty }drr^{2}F(r)=\int_{0}^{\infty }drr^{2}f(r)=\frac{k_{F}^{3}}{%
2}.
\end{equation}%
The above equation states that the integrated orbital moment of both the
regularized and the divergent velocity fields are the same. 
\begin{equation}
G_{i}^{L}=\frac{2m_{e}\eta _{0}k_{F}^{3}}{3\pi ^{2}\hbar }\approx 3.2\times
10^{-3}\left( \frac{k_{F}\mathrm{\mathring{A}}}{1.75}\right) ^{3}.
\end{equation}%
where in the second step we used $\eta _{0}$ for an Al host metal. The
constant $G_{i}^{L}$ plays the role of a $g$-factor, and then the rotational
current contributes by about 0.3\% to the total 4f orbital moment.

\section{\O rsted field generated by the equilibrium currents}

\label{Sec:SI:Ampere}

According to Maxwell's equations the equilibrium charge current vortex
around the rare-earth moment generates a magnetic field Eq. (\ref%
{EqMagnField}):  
\begin{equation*}
\mathbf{B}_{\text{\textrm{\O }}}(\mathbf{R})=-\frac{e\mu _{0}}{4\pi }\int
d^{3}r^{\prime }\left\langle \mathbf{v}(\mathbf{r^{\prime }})\right\rangle
\times \frac{\mathbf{R}-\mathbf{r^{\prime }}}{|\mathbf{R}-\mathbf{r^{\prime }%
}|^{3}}.
\end{equation*}%
The derivation of an analytic expression for general $\mathbf{R}$ is
tedious. However, far from the RE ion, $R=\left\vert \mathbf{R}\right\vert
\gg \langle r\rangle $the Taylor expansion of $\left\vert R\mathbf{\hat{R}}-%
\mathbf{r}\right\vert ^{-3}$ gives 
\begin{equation}
\mathbf{B}_{\text{\O }}(\mathbf{R})=\frac{\mu _{0}}{4\pi }\frac{3\left( 
\mathbf{m_{rc}}\cdot \mathbf{R}\right) \mathbf{R}-R^{2}\mathbf{m_{rc}}}{R^{5}%
},
\end{equation}%
where $\mathbf{m_{rc}}=-\gamma _{0}G_{i}^{L}\mathbf{L}$. The above
expression is the expected result of a field generated by the magnetic
moment $\mathbf{m_{rc}}$ of the current vortex.

On the other hand, the magnetic field at the origin $\mathbf{R}=0$ reads 
\begin{align}
\mathbf{B}_{\text{\textrm{\O }}}(0)& =\frac{e\mu _{0}}{4\pi }%
\int_{0}^{\infty }dr\int_{0}^{\pi }d\theta \sin \theta \int_{0}^{2\pi }d\phi
\left\langle \mathbf{v}(\mathbf{r})\right\rangle \times \mathbf{\hat{r}}, 
\notag \\
& =\frac{\eta _{0}}{2\pi ^{3}\hbar }\frac{e\mu _{0}}{4\pi }\int_{0}^{\infty
}dr\frac{F(r)}{r}\int_{0}^{\pi }d\theta \sin \theta \int_{0}^{2\pi }d\phi
\left( \mathbf{L}\times \mathbf{\hat{r}}\right) \times \mathbf{\hat{r}} 
\notag \\
& =-3.5\times 10^{-7}\frac{e\eta _{0}\mu _{0}}{\hbar a^{6}}\mathbf{L}
\end{align}

\section{Orbital angular momentum density in second quantization}

\label{Sec:SI:orbital} The Pauli equation for an electron wave function $%
\boldsymbol{\psi }\left( \mathbf{r}\right) $ with energy $E$ in a
homogeneous magnetic field $\mathbf{B}$ reads 
\begin{equation}
E\boldsymbol{\psi }\left( \mathbf{r}\right) =\left[ \frac{1}{2m_{e}}\left(
-i\hbar \nabla +e\mathbf{A}\right) ^{2}+\frac{e\hbar }{2m_{e}}\boldsymbol{%
\sigma }\cdot \mathbf{B}\right] \boldsymbol{\psi }\left( \mathbf{r}\right) .
\end{equation}%
In the symmetric gauge $\mathbf{A}=\frac{1}{2}\mathbf{B}\times \mathbf{r}$ 
\begin{align}
E=& \int d^{3}{r}\boldsymbol{\psi }^{\dagger }\left( \mathbf{r}\right) \left[
\frac{1}{2m_{e}}\left( -i\hbar \nabla +\frac{e}{2}\mathbf{B}\times \mathbf{r}%
\right) ^{2}+\frac{e\hbar }{2m_{e}}\boldsymbol{\sigma }\cdot \mathbf{B}%
\right] \boldsymbol{\psi }\left( \mathbf{r}\right)   \notag \\
=& \int d^{3}{r}\boldsymbol{\psi }^{\dagger }\left( \mathbf{r}\right) \left[
-\frac{\hbar ^{2}\nabla ^{2}}{2m_{e}}-ie\hbar \frac{\left( \mathbf{B}\times 
\mathbf{r}\right) \cdot \nabla +\nabla \cdot \left( \mathbf{B}\times \mathbf{%
r}\right) }{4m_{e}}\right.   \notag \\
& \left. +\frac{e\hbar }{2m_{e}}\boldsymbol{\sigma }\cdot \mathbf{B}+%
\mathcal{O}\left( B^{2}\right) \right] \boldsymbol{\psi }\left( \mathbf{r}%
\right) .
\end{align}%
Then, the energy of the Zeeman coupling $E_{Z}$ is 
\begin{equation}
E_{Z}=\frac{e}{2m_{e}}\left( \mathbf{l}_{T}+2\mathbf{s}_{T}\right) \cdot 
\mathbf{B},
\end{equation}%
where the factor 2 is the single-electron orbital $g$-factor. In terms of
the total spin $\mathbf{s}_{T}$ and orbital $\mathbf{l}_{T}$ angular momenta 
\begin{align}
\mathbf{s}_{T}& =\frac{\hbar }{2}\int d^{3}{r}\boldsymbol{\psi }^{\dagger
}\left( \mathbf{r}\right) \boldsymbol{\sigma }\boldsymbol{\psi }\left( 
\mathbf{r}\right) , \\
\mathbf{l}_{T}& =\frac{i\hbar }{2}\int d^{3}{r}\boldsymbol{\psi }^{\dagger
}\left( \mathbf{r}\right) \left( \nabla \times \mathbf{r}-\mathbf{r}\times
\nabla \right) \boldsymbol{\psi }\left( \mathbf{r}\right) .
\end{align}%
Substituting 
\begin{align}
\boldsymbol{\psi }(\mathbf{r})& =\frac{1}{\sqrt{\Omega }}\sum_{\mathbf{p}%
,\alpha }e^{i\mathbf{p}\cdot \mathbf{r}}\boldsymbol{\chi _{\alpha }}a_{%
\mathbf{p}\alpha }, \\
\boldsymbol{\psi }^{\dagger }(\mathbf{r})& =\frac{1}{\sqrt{\Omega }}\sum_{%
\mathbf{q},\beta }e^{-i\mathbf{q}\cdot \mathbf{r}}\boldsymbol{\chi _{\beta }}%
^{\dagger }a_{\mathbf{q}\beta }^{\dagger },
\end{align}%
where the spinors $\boldsymbol{\chi _{\uparrow }}$ and $\boldsymbol{\chi
_{\downarrow }}$ are the basis of $\sigma _{z}$. The second quantized
version of $\mathbf{s}(\mathbf{r})$ and $\mathbf{l}(\mathbf{r})$, the local
densities of spin and orbital momentum (relative to the origin),
respectively:%
\begin{align}
\mathbf{s}_{T}& =\int \mathbf{s}(\mathbf{r})d^{3}{r}, \\
\mathbf{s}(\mathbf{r})& =\frac{\hbar }{2}\frac{1}{\Omega }\sum_{\mathbf{pq}%
\alpha \beta }e^{i\left( \mathbf{p}-\mathbf{q}\right) \cdot \mathbf{r}}a_{%
\mathbf{q}\beta }^{\dagger }\boldsymbol{\sigma }_{\alpha \beta }a_{\mathbf{p}%
\alpha },
\end{align}%
and 
\begin{equation}
\mathbf{l}_{T}=\frac{i\hbar }{2}\int d^{3}{r}\boldsymbol{\psi }^{\dagger
}\left( \mathbf{r}\right) \left( \nabla \times \mathbf{r}-\mathbf{r}\times
\nabla \right) \boldsymbol{\psi }\left( \mathbf{r}\right) =\int \mathbf{l}(%
\mathbf{r})d^{3}{r},
\end{equation}%
with 
\begin{align}
\mathbf{l}(\mathbf{r})& =\frac{\hbar }{\Omega }\mathbf{r}\times \sum_{%
\mathbf{pq}\gamma }e^{i\left( \mathbf{p}-\mathbf{q}\right) \cdot \mathbf{r}}%
\mathbf{p}a_{\mathbf{q}\gamma }^{\dagger }a_{\mathbf{p}\gamma }  \notag \\
& =\frac{\hbar }{\Omega }\mathbf{r}\times \sum_{\mathbf{pq}\gamma
}e^{i\left( \mathbf{p}-\mathbf{q}\right) \cdot \mathbf{r}}\frac{\mathbf{p}+%
\mathbf{q}}{2}a_{\mathbf{q}\gamma }^{\dagger }a_{\mathbf{p}\gamma }.
\end{align}%
Note that $\mathbf{l}_{T}=m_{e}\mathbf{r}_{op}\times \mathbf{v}_{op}$, where 
$\mathbf{r}_{op}$ and $\mathbf{v}_{op}$ are the position and velocity
operators, respectively.

\twocolumngrid

\end{document}